\documentclass{nature}

\usepackage{xcolor}

\usepackage{graphicx}
\makeatletter
\let\saved@includegraphics\includegraphics
\AtBeginDocument{\let\includegraphics\saved@includegraphics}
\renewenvironment*{figure}{\@float{figure}}{\end@float}
\makeatother

\title{Cracking the Quantum Advantage threshold for Gaussian Boson Sampling}

\author{A. S. Popova$^{1,2,*}$ \& A.N. Rubtsov$^{1,3}$}

\begin{document}

\maketitle

\begin{affiliations}
 \item{Russian Quantum Center, Bolshoy Bulvar 30, bld. 1, Skolkovo, Moscow 121205, Russia}
 \item{Moscow Institute of Physics and Technology, Dolgoprudny, Moscow Region 141700, Russia }
  \item{Department of Physics, Lomonosov Moscow State University, Leninskie gory 1, Moscow 119991, Russia}
\end{affiliations}

\begin{abstract}
Scientists in quantum technology aspire to quantum advantage: a computational result unattainable with classical computers. Gaussian boson sampling experiment has been already claimed to achieve this goal. In this setup squeezed light states interfere in a mid-sized linear optical network, where multi-photon collisions take place. The exact simulation of the counting statistics of  $n$ threshold detectors is far beyond the possibilities of modern supercomputers once $n$ exceeds $100$. Here we challenge quantum advantage for a mid-sized Gaussian boson sampling setup and propose the approximate algorithm to obtain the probability of any specific measurement outcome. For an 70-mode device on a laptop, our approximation achieves accuracy competitive with the experimental one.

\end{abstract}

\section{Introduction}

\begin{figure}
    \centering
	\includegraphics[width=0.73\linewidth]{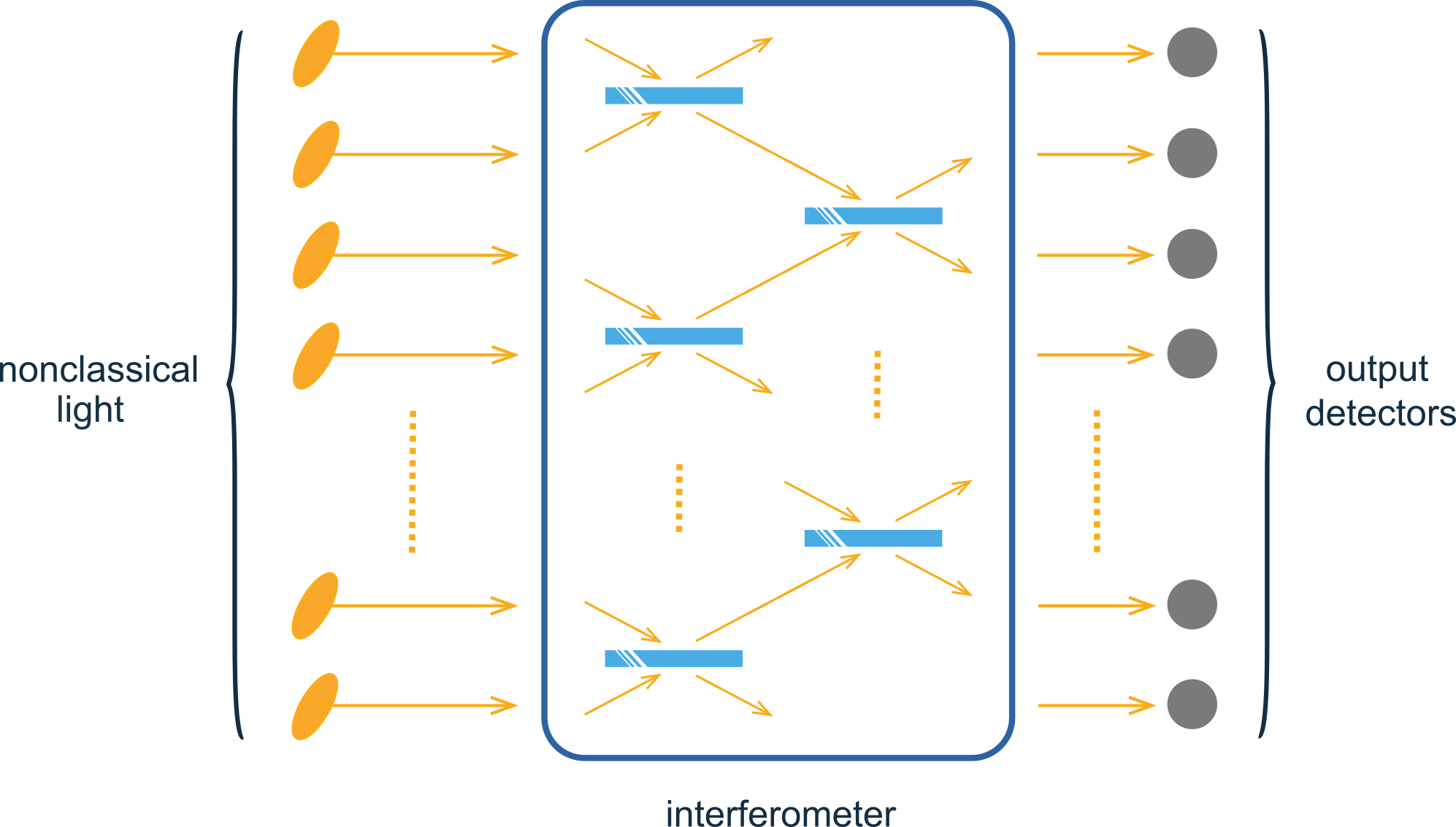}

	\caption{\textit{Boson sampler: non-classical light states
transferring through a linear interferometer are measured with output detectors.}}
	\label{fig:1}
\end{figure}

Quantum computing devices process information in a way unachievable for classical machines. Scientists have already developed a number of quantum processors based on different technology, including superconducting qubits~\cite{arute2019quantum,kjaergaard2020superconducting}, trapped ions~\cite{zhang2017observation,bruzewicz2019trapped}, neutral atoms~\cite{bernien2017probing,henriet2020quantum} and photonic hardware~\cite{qiang2021implementing, arrazola2021quantum}. Here, we appeal to a particular class of quantum devices - boson samplers. These devices operate with non-classical light sources which are mixed by a linear optical network, as depicted in~Figure~\ref{fig:1}. Although these machines cannot perform universal quantum computation, they have the potential of solving practical tasks, such as calculation of the vibronic spectrum of a molecule~\cite{huh2015boson,huh2017vibronic}, molecular docking~\cite{banchi2020molecular}, statistical modelling~\cite{jahangiri2020point}, counting the number of perfect matching in a graph~\cite{bradler2018gaussian} and machine learning~\cite{schuld2020measuring, banchi2020training}.

Boson samplers became a hot topic because of the quantum advantage claimed for this technology. If someone wished to calculate the probability that photons came in 60 detectors, it would take millions of years of any supercomputer, even with randomly chosen circuit parameters. This deceptively simple setup was proposed in the seminal paper~\cite{aaronson2011computational} where single-photon emitters were utilized as non-classical light sources. 

The first experimental implementations of boson samplers followed the original proposal~\cite{aaronson2011computational} but using single-photon sources limits the practical scalability of such devices. Among the recent experiments are the integrated 21-mode optical circuit with 5 input photons~\cite{carolan2014experimental} and the 60-mode interferometer with 20 input photons~\cite{wang2019boson}. On the other hand, the quantum advantage threshold is on about 50 injected photons for boson sampling, this number of synchronous single-photon sources is currently far beyond the best experimental achievements.

The complications caused by the single-photon sources fostered the Gaussian boson sampling (GBS) proposal - a linear-optical scheme with single-mode squeezed light on inputs~\cite{hamilton2017gaussian}. The output wave-function, up to a normalizing constant, equals (see Methods section).
\begin{equation}
|X\rangle=  e^{-\sum_{ii'} X_{ii'} a^\dag_i a^\dag_{i'}}|0\rangle ,
\label{eqn:wf}
\end{equation}
 where $a_i^\dag$ is the creation operator for the $i$th output channel, $|0\rangle$ is the vacuum state, and $X_{ii'}$ is a complex-valued symmetric matrix determined by the interferometer unitary matrix and squeezing parameters of the input light. To obtain the exact probability of a specific output, a classical algorithm have to solve a hard problem: find the projections of the wave-function~(\ref{eqn:wf}) onto a measurement basis, which requires exponentially large number of elementary operations.

Zhong et al. claimed the quantum advantage for the GBS experiment with 50 input squeezed states mixed into a 100-mode~\cite{zhong2020quantum} and 144-mode~\cite{zhong2021phase} interferometers. The detectors of these schemes click in presence of one or more photons in output channels (threshold detection), but cannot count photons in each channel. The number of clicked detectors fluctuates to $n=113$ for the 144-mode set-up~\cite{zhong2021phase} in accordance with a non-classical probability distribution. In comparison, the Sunway TaihuLight supercomputer spent two days finding a probability of a single $n = 50$ measurement outcome~\cite{li2021benchmarking}.But it might need less time if the supercomputer had simulated the realistic GBS device with reported imperfections. And how much less the computational time can be for the GBS simulation is the question currently under active discussion.

 The direct comparison of the experimental data and the classical simulation faces difficulties. The main theoretical efforts were focused on one or two types of potential GBS imperfections, such as  photon losses~\cite{aaronson2016bosonsampling, oszmaniec2018classical, garcia2019simulating, brod2020classical, oh2021classical,deshpande2022quantum}, noise in an interferometer~\cite{kalai2014gaussian,arkhipov2015bosonsampling,oh2021classical} and distinguishability of input states~\cite{rohde2012error, renema2020simulability}. To take into account all these imperfections together for the real experiment, the non-classicality test~\cite{qi2020regimes} was proposed as a necessary condition for the non-trivial output distribution (which was passed by the recent experiment of Zhong et al.~\cite{zhong2020quantum}). However, this test does not include the presence of multi-photon collisions as an imperfection. 

Although the exact GBS simulation requires exponential time, the classical algorithms moved the quantum advantage threshold to $n=100$ clicked detectors~\cite{bulmer2021boundary}. This fact mainly connects with two sources of decoherence: multi-photon collisions and photon losses.  First of all, the Boson sampling complexity (as the Gaussian boson sampling) was proven for the case when the number of the injected photons was significantly less than the number of modes $n_{ph} \ll m$, for example, $n_{ph}\sim\sqrt{m}$. The recent Gaussian boson sampling experiments have the average number of photons is proportional to the number of modes $ \overline{n_{ph}} \sim {m}$, we can conclude that multi-photon collisions take place. Strictly speaking, there is no guarantee that an effective classical method cannot be created for this regime. Secondly, the overall transmission rate in the experiment is around $0.5$~\cite{zhong2021phase}, in other words, half of the injected photons are lost. Nevertheless, exactly this regime lies on the frontier of possibilities of classical algorithms: the losses level is significantly larger for rigorously claiming the GBS hardness and too low for the opposite statement. The actual scheme was simulated most closely with chain-rule simulation and Metropolis independence sampling~\cite{bulmer2021boundary} -- the authors created the exact classical simulation, which shifted the quantum advantage threshold for the GBS experiment to $n=100$ clicked detectors in the presence of multi-photon collisions.

  
In this work, we develop a new polynomial-time approximate algorithm to emulate GBS devices with threshold detection and multi-photon collisions. To calculate outcome probabilities, we introduce a series of approximations, treating the GBS emulation as a reversed problem of statistical physics. We estimate absolute probabilities of random samples and compare them to the exact values. We do not include the the other sources of decoherence, such as photon losses, noise in an interferometer and distinguishability of input states. 

 \section{Results} 
 
We describe the core of our method based on the example of the specific mid-sized GBS problem. After this, we show the main emulation results of larger GBS setups and the numerical accuracy of our approximate approach.

\begin{figure}
    \centering
	\includegraphics[width=0.7\linewidth]{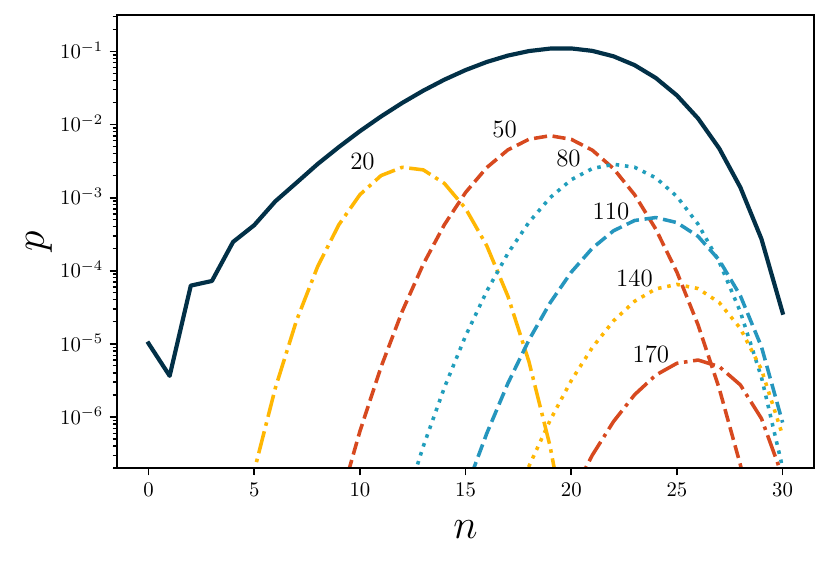}
	\caption{\textit{The total probability $p(n)$ to detect photons in $n$ channels  of the 30-mode interferometer with 15 input single-mode squeezed states (solid line) and the contributions $p_{k}(n)$ to  $p(n)$ from some of the  $k$-sectors (dashed lines), where $k$ is a number of photons injected into the scheme. The data for $k=20,50,80,110,140$, and $k=170$ are shown; the squeezing parameter is $r = 1.4$ for each input, the interferometer’s matrix is a random unitary.}}
	\label{fig:2}
\end{figure}

  \textsl{Calculation of $p(n)$}. Our results rely on the consideration of the probability $p(n)$ of photon detection in any $n$ outputs of an $m$-mode interferometer. For the scheme is equipped with $m$ threshold detectors, $p(n)$ is a sum of probabilities over all possible $n$-combinations of $m$ outputs. Figure~\ref{fig:2} shows $p(n)$ computed with a numerically exact method for a mid-sized GBS device - a 30-mode interferometer with 15 input single-mode squeezed vacuum states. 
 
  \textsl{Calculation of sectors $p_k(n)$}. Although the number of photons coming in detectors is not resolved in the recent GBS experiment, we use it in the theoretical description. The photon-number operator commutes with observables, and therefore $p(n)$ can be presented as a sum over sectors corresponding to different photon numbers $k$: $p(n)=\sum_k p_{k}(n)$ (see Methods). The sectors contribute to the total probability differently as depicted in Figure~\ref{fig:2}. In particular, large $k$ have dominant contributions to the $p(n)$ for $n\approx m$ point, because the maxima of $p_k(n)$ shift to the right with increasing $k$. The point $n=m$ is special. It corresponds to a single possible measurement outcome when all detectors clicked so that
\begin{equation}
    p_{k}(m)=\frac{\langle X_{k}|\Pi_{1,2 ..., m}| X_{k}\rangle }{\langle X|X\rangle}
\end{equation}
where $\Pi_{1,2, ..., m}$ is a projector to the subspace in which photons are present in all modes $1,2, ..., m$, and $| X_{k}\rangle$ is the projection of $| X\rangle$ to the sector $k$.  Further we will provide an approximate procedure for $p(n)$  based on the momentum expansion for the $p_{k}(n)$ in each sector. Later we also will show that this procedure is applicable not only for the $n=m$ case but also can be extended for any specific measurement outcome with $n<m$.

 \textsl{Approximation scheme}. Our approach is based on two observations: 1) the dependencies $p_k(n)$ are smooth curves; 2) the low-order moments $\sum_n n^j p_k(n)$ calculation run polynomial time.  These observations suggest the existence of an approximate method to compute $p_k(n)$ via low-order moments. Although the general inverse moment problem is known to be ill-posed~\cite{john2007techniques}, our problem is defined better: each sector is characterized by a single-peak (unimodal) distribution $p_k(n)$ and sectors with the largest contribution have maxima close to the $n=m$ point.  
We approximate the distribution $p_k(n)$ in each sector with an exponent of the polynomial, $p_{k}(n)  \approx e^{-f_j(n)}$, where $j$ is the order of polynomial; for example for $j = 2$, $p_k(n)$ is approximated by a Gaussian distribution. We also outline a way to calculate the set of moments up to the 4th order using $m^5$ elementary operations in the Methods section: each next order $j$ requires an extra power of $m$. 

Our approach has three steps that are described in Methods section: in the first step, we calculate the partition functions of an entire system and any measurement outcome; in the second step, we compute moments up to $j$-order;  in the third step, we approximate $p_{k}(n)$ and obtain the sum over $k$ sectors the last point $n=m$ -- the probability $p(n)$ of a measurement outcome. 

\begin{figure}
    \centering
		\begin{minipage}[h]{1\linewidth}
		\includegraphics[width=1\linewidth]{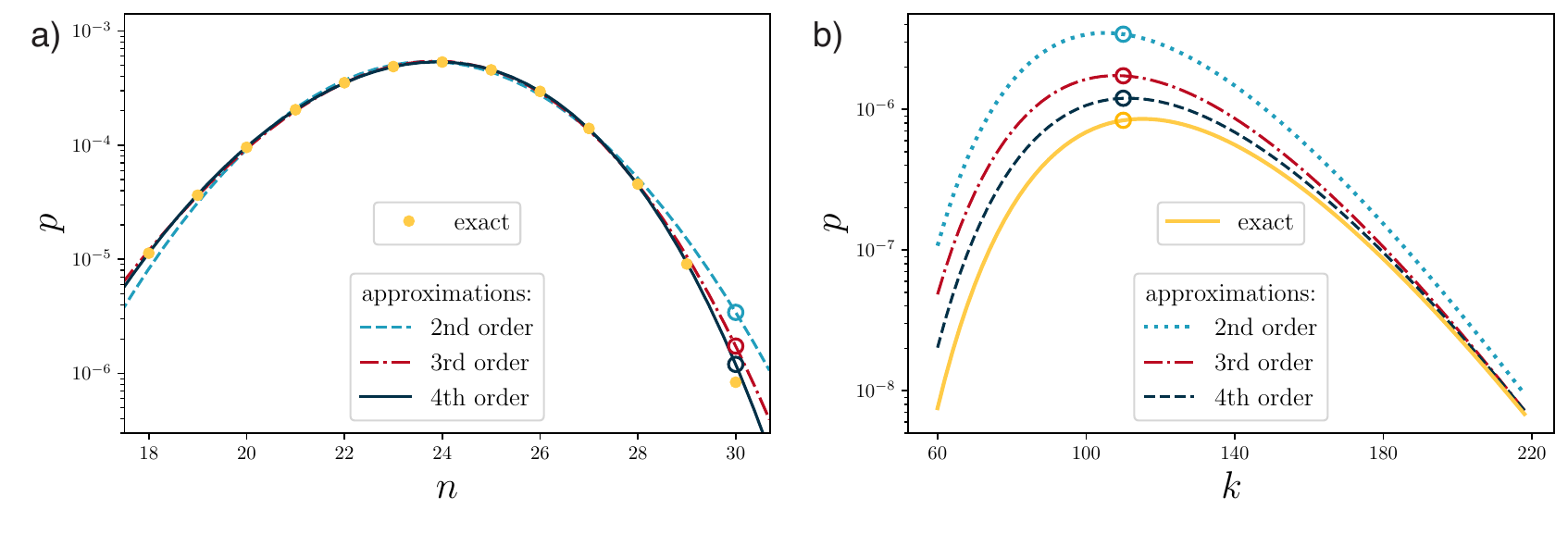} 
		
	\end{minipage}

	\caption{\textit{ a) The contribution $p_{k}(n)$ to the total distribution $p(n)$ of $k = 110$ photons in the 30-mode GBS setup; b) Comparison of the approximations with the various number of moments for the last $n=30$ point probability  (i.e. for the case when photons come into all detectors).}}
	\label{fig:3}
\end{figure}

Figure~\ref{fig:3} shows the different orders of approximation for the probability $p(n)$ when photons come into all $30$ detectors of the benchmark GBS device. The left panel of Figure~\ref{fig:3} depicts the comparison of the exact probabilities to our solution for the $k = 110$ sector. As we can see, the approximation performance is improved by increasing the number of  moments accounted.  This picture is almost the same for any sector $k$ of the benchmark setup, as shown in the right panel of Figure~\ref{fig:3}. The 2nd-, 3rd-, and 4th-order approximations for $p(n)$ and $n = 30$ deviate from the exact value by $77\%$, $49\%$,  and  $28\%$ respectively.

 \textsl{Probability of any specific sample}. Let us now turn to the case of a generic measurement outcome, when photons came only in some~$i_1 ... i_{n}$~detectors, $n<m$. The probability of this event is given by
\begin{equation}
p(n) = \frac{\langle X'|\Pi_{i_1... i_{n}}| X'\rangle }{\langle X'|X'\rangle} \frac{\langle X'|X'\rangle}{\langle X|X\rangle},     
\label{eqn:p(n)}
\end{equation}

where $|X'\rangle=e^{-\sum_{i,i'=i_1...i_{n} } X_{ii'} a^\dag_i a^\dag_{i'}}|0\rangle$. The structure of the first factor in the r.h.s. is the same as for $p(m)$: only the matrix $X'$ is a submatrix built from the rows and columns  $i_1 ... i_{n}$ of the specific measurement outcome. The second factor can be calculated in a polynomial time (see Methods section). Thus we can apply our approach directly and estimate the probability that photons came in the $n$ desired detectors.

\begin{figure}
    \centering
    \begin{minipage}[h]{1\linewidth}
	\includegraphics[width=0.55\linewidth]{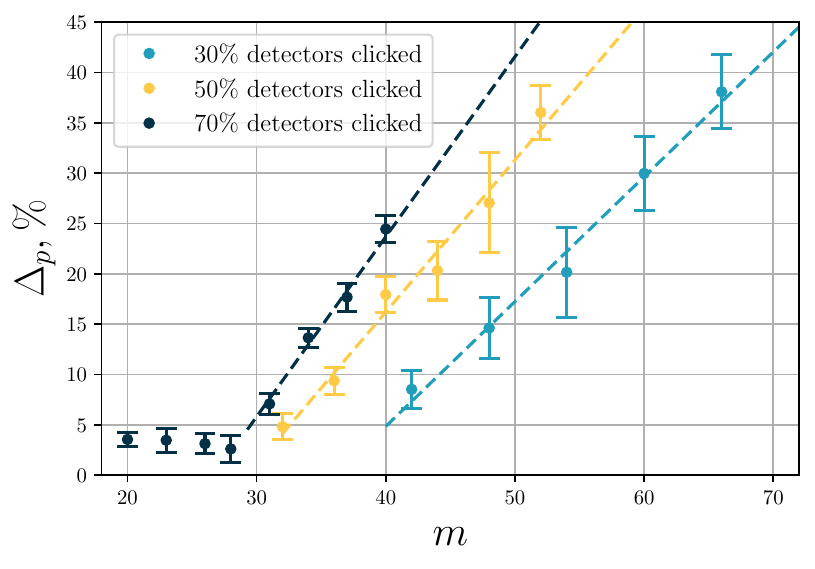}
	\includegraphics[width=0.45\linewidth]{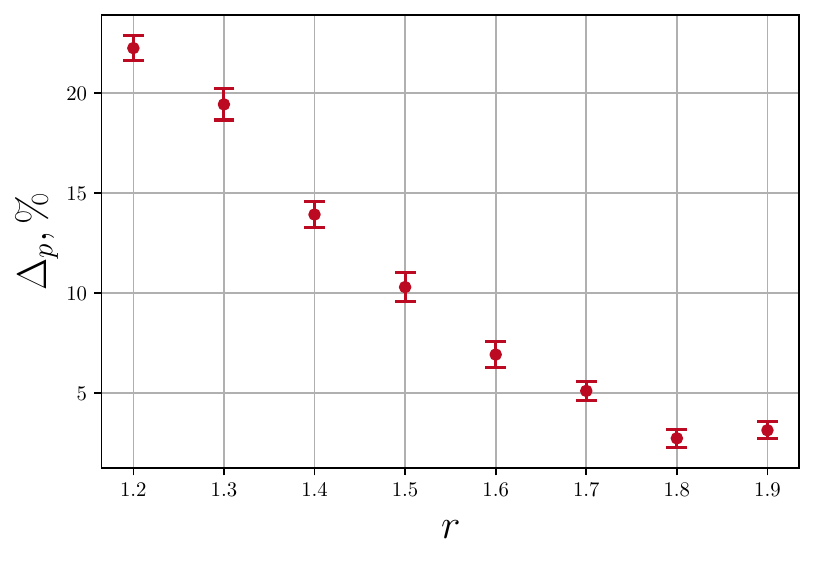}
	\end{minipage}
	\caption{\textit{The relative deviation $\Delta_p$ between the exact value of the probability and an estimated one with the 4th-order scheme as a function of the number of modes $m$ (left panel, $r = 1.6$) and squeezing parameter $r$ (right panel, 20 clicked detectors in the 30-mode interferometer);  bars show the standard deviation of the approximate results, dashed lines are to guide the eye.}}
	
	\label{fig:4}
\end{figure}

   \textsl{Numerical accuracy}. To understand how our method performs in general, we compared the forth-order approximation for a set of GBS setups and different number of the detectors clicked. The results are presented in Figure~\ref{fig:4}, left panel: it shows the relative deviation of probabilities and its variance, each point is an average over 100 random measurement outcomes. For example,  the relative deviation of probabilities for $n=18$ clicked detectors was compared for $m= 54, 36, 26-$mode random interferometers~(which is $30\%$, $50\%$ and $70\%$ of clicked detectors respectively in Figure~\ref{fig:4}).  Variances shown in Figure~\ref{fig:4} signals that no estimated probability differs dramatically from its exact value.Notably, the accuracy of our method increases with the growth of squeezing parameter (depicted in the right panel of Figure~\ref{fig:4}), where GBS scheme has a half of filled inputs), which contrasts to previous strategies for classical GBS simulation~\cite{qi2020regimes}.
    
    We estimate the domain of applicability of our scheme from the requirement that the deviation should be less than about $80\%$. 
    As Figure~\ref{fig:4} shows the relative deviation grows linearly with $m$ starting from some point for similar proportion of clicked detectors. We extrapolate this linear behaviour to higher $m$ there no exact reference data is available. This leads to the conclusion that our 4-th order scheme can at least evaluate probabilities for 50 clicked detectors of a 70-mode interferometer.

 \section{Discussion}
 
We proposed a series of approximations to evaluate the probabilities of any measurement outcome  of the Gaussian boson sampling experiment. We empirically established a good convergence of our method to the exact result for an setup with a random interferometer matrix, multi-photon collisions and threshold detectors. The 4-th order algorithm is able to evaluate probabilities of the $n = 50$ clicked detectors outcomes in a 70-mode GBS scheme on a laptop. 

We recall the main attributes of Gaussian boson sampling that allowed us for a successful emulation. The first is the exact polynomial formulas for the moments~$p(n)$, and the second is the distribution~$p_k(n)$ in the sector with $k$~photons obeying unimodal distribution. Further improvement of the accuracy can be achieved by 1)~using a more optimal guess function for the $p_k(n)$; 2)~including higher orders of approximation~$j$, but this would require longer calculation time  $ O(n^{j+1})$. 

It is important to stress that we emulate an GBS setup and compare our method with the exact data in presence of multi-photon collisions. We neglect other imperfections, such as photon losses, noise in an interferometer and distinguishability of inputs, which usually present in realistic set-ups. When we compare our simulation with the experiment, which has at least 50$\%$ photon losses, we assume the experimental output probability distribution significantly differs from the output of the lossless scheme. Because we compare our approximate method with the exact lossless computation, we similarly consider  the experimental results. In this perspective, the relative deviation between the experiment and the exact data is comparable with our results. 

It is worth drawing attention to the fact that we calculate absolute probabilities of randomly generated samples and we do not provide samples from the non-classical distribution. However, we are guided by a simple conjecture: if one can evaluate the absolute probabilities in polynomial time, it is possible to create a fast sampling algorithm -- for example, using a Markov chain Monte-Carlo method~\cite{neville2017classical} with our 4-th order approximation as the proposal distribution. So-called ``spoofers'' are also able to produce samples from a non-classical probability distribution, for example, the Boltzmann machine and greedy algorithm, but it includes up to the 2-order marginals~\cite{villalonga2021efficient}: the computational time of the method grows exponentially with k to reproduce k-order correlations. Here we were more focused on the approximate procedure of the evaluation of random samples because our current algorithm can be also used for the Bayesian test~\cite{bentivegna2015bayesian} for the real GBS devices validation. Creating an effective sampling algorithm is the next step in our work.

We believe that our approach might give better performance for lossy and noisy devices because their simulation typically requires fewer resources~\cite{brod2020classical,qi2020regimes,oh2021classical}. In further work, we will focus on the GBS devices with other sources of decoherence to compare the results of our method with the experimental data appropriately.

\section{Methods}

Here we describe our approximate polynomial-time algorithm for an idealized GBS setup with threshold detectors in detail.  To calculate the probability of a detector’s event this method includes three steps:
1)~defining the partition function and the sector distribution of the multimode Gaussian state of the entire system and a specific measurement outcome, 
2)~moments’ calculation, and 3)~approximation the sector distribution via its moments. 

\textsl{Model}. Gaussian boson sampling devices consist of inputs filled with a squeezed light and an interferometer $U$ equipped by threshold detectors. A single-mode squeezed vacuum  state is described by the wave function $|x\rangle=e^{-x a^\dag a^\dag}|0\rangle=\sum_l \frac{(-x)^l \sqrt{(2l)!}}{l!} |2 l\rangle$  in the Fock basis~\cite{walls2007quantum}. Summing the non-normalized series yields $\langle x|x\rangle=\sum_l \frac{|x|^{2 l} (2l)!}{(l!)^2}=\frac{1}{\sqrt{1-4 |x|^2}}$. Thus the multimode Gaussian state at the output of the GBS device is $|X\rangle=e^{-X_{ii'} a^\dag_i a^\dag_{i'}}|0\rangle$, where $X_{ii'} = \frac{1}{2}\sum_j U_{ji} U_{ji'} x_j$ is a symmetric complex-valued matrix (presented  using  the Autonne–Takagi factorization~\cite{horn2012matrix}), $x_j = \tanh{(r_j) }$, where $r_j$  is squeezing parameter in $j$-mode, and $i,i'$ is the number of output and input channels respectively. The operator of observable is $\Pi_j = I - |0_j \rangle \langle 0_j|$ that corresponds a click in $j$-detector.

\textsl{Partition function}. We consider threshold detection as certain Grand-canonical statistical ensemble with the $Z_X$ partition function, which is the sum of the weights of all possible measurement outcomes. The partition function of the entire system equals to 
\begin{equation}
 Z_X\equiv \langle X|X\rangle=\frac{1}{\sqrt{\det({I}-X^\dag X)}},  
\end{equation}
where $X^\dag$ is a Hermitian conjugate of the matrix $X$. On the other hand, $Z_X$  is equal to the sum over the sectors with different numbers of photons in the system:  $Z_X^k=\langle X| \delta(\sum_i a^\dag_i a_i -k) |X \rangle$. After the Kronecker symbol $\delta$ is expanded in the Fourier series, we obtain $Z_X^k= \int Z_X^\nu e^{-{\bf{i}} k \nu} d\nu$ with $Z_X^\nu=\langle X| e^{{\bf{i}} \nu \sum_i a^\dag_i a_i} |X \rangle$.   In practice, the partial sum  $Z_X^k$  (formed by the first $k$ terms)  is a sum over a discrete set of $\nu$. The values of $Z_X^{\nu}$ are obtained by performing the Fourier expansion of the operator exponents in the $X$ basis, that gives
\begin{equation}
Z_X^\nu=\frac{1}{\sqrt{\det({I}- e^{{\bf{i}} \nu} X^\dag X)}}.
\end{equation}

\textsl{Probability of any sample}. Next, we show how to calculate the probability of any specific measurement outcome, when photons came in a random subset of $m$ detectors $\{ i_1, i_2, i_3, ..., i_m \}$ ( if $a$-detector clicked $i_a = 0$ and $i_a=1$ otherwise for any $a$). For this goal we consider a sub-ensemble in which photons were detected in $n$ channels labelled $\{ i_1, .., i_{n}\}$, whereas other channels definitely remain in the vacuum state. Thus, the overall Gaussian state $|X\rangle$ can be projected to the corresponding subspace by putting to zero the creation operators  $a^{\dagger}_{q}$ with $q \neq \{i_1 ,..., i_{n}\}$. The projected state $|X'\rangle$ is a Gaussian state also, but it is characterized by a submatrix of $X_{ii'}$ with  $i,i'= \{ i_1 ... i_{n}\}$. In this way, the probability of a specific measurement outcome can be calculated using Equation~(\ref{eqn:p(n)}): it is the last point of the probability distribution via the sub-ensemble,  multiplied by the weight  $Z_{{X'}}/Z_{X}$. This is an exact computation of the GBS problem with threshold detection that requires the computation of $2^{n}$ determinants.

\textsl{Moments calculation}. We can also calculate moments under our representation of the GBS problem. For example, the average number of clicked detectors of a GBS device is equal to
 \begin{equation}
 \bar{n} = \sum_{l} \left(1- \frac{Z_{X_l}}{Z_X}\right)
 \end{equation}
where $X_{l}$ equals $X$ with one vanished $l$-column and row; the first moment of a sub-ensemble requires adding the sum over $k$-sectors $Z_{X_l}^{k}$.  The second- and higher-order moments can be expressed with the same procedure with second- and higher-order minors. 
Thus, the probability of any specific outcome is the last point of distribution in a sub-ensemble and the exact calculation of it requires exponential resources.

\textsl{Approximation method}. We use our representation of the GBS problem to propose an approximate polynomial-time procedure. It is based on the estimation of the last point in the sub-ensemble as a sum over approximate sector distributions. We estimate the sector distribution via its moments so that we need to compute determinants of multiple similar matrices. Using the Sherman-Morrison approach~\cite{sherman1950adjustment} enabled us to accelerate the computation: for the 4th- and lower-order moments in all sectors our algorithm runs using $\propto n^5$ operations. 

 The last step of our approach is solving an inverse moment problem for sectors. We iteratively find parameters of the approximate probability distribution $p_{k}(n)  \approx e^{-f_j(n)}$ in each sector so that moments of this distribution equals the true moments. For example, the second-order scheme has a distribution
 \begin{equation}
 p_{k}(n)  \approx  c_0 e^{-(n - c_1)^2/(2 c_2^2)} 
 \end{equation}
 the calculated parameters $c_0, c_1, c_2$ give the first $\bar{n}$ and second moment accurately. In fact, the true distribution is not a Gaussian one and higher moments reproduce its shape better, which is shown in Figure~\ref{fig:3}(b). The 4th-order scheme evaluates output probabilities with quite high precision and might be improved by optimization of the distribution's form.  The most greedy part of our method is the 4th-order moments computation for the sectors, which requires extra 30 GB of memory for 50 clicks in our implementation and can be accelerated by a more memory-efficient algorithm.


\bibliography{gbs_ref}

\begin{addendum}
 \item This work was carried out in the framework of the Russian Quantum Technologies Roadmap. 
 
 \item[Code availability] https://github.com/stacy8popova/PyGBSThr 
 
 \item[Data availability] https://doi.org/10.13140/RG.2.2.19178.24004

 \item[Author Contribution] A.N. Rubtsov proposed the approximate scheme and both named authors equally contributed to conducting the further research and preparation of this manuscript.

 \item[Competing Interests] The authors declare that they have no competing financial interests.

 \item[Correspondence] Correspondence and requests for materials 
should be addressed to Anastasia Popova (email:~popova.as@phystech.edu).
\end{addendum}

\end{document}